%% file: Main_v3.tex
\documentclass[nofootinbib]{article}

\usepackage{amssymb}

\usepackage{amsmath}

\usepackage{bm}

\usepackage{jcappub}

\include{mydefs}

\newcommand{\lf}{{\sc lat}field{\sc 2}}

\title{A numerical relativity scheme for cosmological simulations}

\author{David Daverio$^{1,2,3}$} 
\emailAdd{dd415@damtp.cam.ac.uk}
\author{Yves Dirian$^1$}
\emailAdd{yves.dirian@unige.ch}
\author{Ermis Mitsou$^{1,4,5}$}
\emailAdd{ermitsou@physik.uzh.ch}

\affiliation{$^1$Department of Theoretical Physics and Center for Astroparticle Physics, University of Geneva, 24 quai Ansermet, CH--1211 Gen\`eve 4, Switzerland}
\affiliation{$^2$African Institute for Mathematical Sciences, 6 Melrose Road, Muizenberg, 7945, South Africa}
\affiliation{$^3$Centre for Theoretical Cosmology, Department of Applied Mathematics and Theoretical Physics, Wilberforce Road, Cambridge CB3 0WA, United  Kingdom}
\affiliation{$^4$Physics Department and Institute for Strings, Cosmology and Astroparticle Physics, Columbia University, New York, NY 10027, USA}
\affiliation{$^5$Center for Theoretical Astrophysics and Cosmology, Institute for Computational Science, University of Zurich, CH--8057 Z\"urich, Switzerland}

\abstract{Cosmological simulations involving the fully covariant gravitational dynamics may prove relevant in understanding relativistic/non-linear features and, therefore, in taking better advantage of the upcoming large scale structure survey data. We propose a new 3+1 integration scheme for General Relativity in the case where the matter sector contains a minimally-coupled perfect fluid field. The original feature is that we completely eliminate the fluid components through the constraint equations, thus remaining with a set of unconstrained evolution equations for the rest of the fields. This procedure does not constrain the lapse function and shift vector, so it holds in arbitrary gauge and also works for arbitrary equation of state. An important advantage of this scheme is that it allows one to define and pass an adaptation of the robustness test to the cosmological context, at least in the case of pressureless perfect fluid matter, which is the relevant one for late-time cosmology.}

\begin{document}

\maketitle

\flushbottom

\section{Introduction}

The upcoming large scale structure observations will provide a powerful probe of relativistic/non-linear effects in cosmology, such as those encountered in alternative descriptions of the dark sector, backreaction, matter-radiation interactions, neutrinos, cosmic defects, the inflationary phase, etc. This motivated some very recent developments in going beyond the standard simulation techniques that are the linear-Boltzmann and Newtonian $N$-body approaches. These include the first relativistic $N$-body code \cite{ADDK1, ADK, ADDK2, ADDK3}, where the gravitational field is treated through an appropriately truncated second-order perturbation theory \cite{GW2, GW, ADDK1}, Newtonian $N$-body simulations that include linearized radiation \cite{FTRCKW, BRTLFH} or a more sophisticated estimate of the scale factor \cite{RDBSC}, and a Boltzmann method for describing non-linear effects of massive neutrinos \cite{BD}. At the same time, fully non-linear simulations of General Relativity (GR) have been performed to study aspects of both the early and late universe, using well-established schemes of numerical relativity (NR, \cite{Lehner, Gourgoulhon, Font, Alcubierre, BBCPL, Shibata} for reviews). There are simulations of the inflationary epoch in $1+1$ \cite{WJPALL, WJAP, JWAP, BJPA} and $3+1$ \cite{EKLS, EKSYZ, CLDNFFP} dimensions, while for late-time cosmology there are $1+1$ spherical collapse simulations \cite{RCCF, RFCC} and $3+1$ simulations involving a pressureless perfect fluid field \cite{GMS, MGS, GMS2, BB, Bentivegna, MLP} (see also \cite{Hern}). The latter are mainly motivated by the issue of cosmic backreaction which, until recently, was addressed numerically only through lattice configurations \cite{CF, CRT, YANT, BL, BK, BL2, BK2}. Finally, there is also a renewal of interest in $3+1$ NR $N$-body simulations of the collapse dynamics \cite{YHO}, after the earlier work of Shibata \cite{Shibata1, Shibata2} and the even earlier, but lower-dimensional, works of Shapiro and Teukolsky using both $N$-body and Boltzmann approaches \cite{ST1, ST2, ST3, ST4, RST, ST5} (see also \cite{Olabarrieta}).

Our interest here lies in $3+1$ NR applied to cosmology. A first observation is that NR has been mostly developed to deal with strong gravity phenomena at astrophysical scales, in which case the scenarios of interest involve localized (if not absent) matter. This is in sharp contrast with cosmology, where gravity is weak and matter is totally delocalized, so there are novel features that need to be taken into account. One such feature is that hydrodynamical matter fields generically contain growing modes around the FLRW background, already at the linear (analytic) level. The prototypical example is the linearly gauge-invariant density contrast $\de^* := \( \ro - \bar{\ro}\) / \bar{\ro} \, + \dots$, which appears in the Poisson equation for the Bardeen potential $\pa^2 \Phi \sim a^2 \bar{\ro}\,\de^*$. For instance, in the case of a pressureless perfect fluid, $\de^*$ contains a growing mode proportional to the scale factor $a$, corresponding to the growth of structure due to gravitational attraction. At $\de^* \sim 1$ perturbation theory breaks down and the fully non-linear matter dynamics must be taken into account in order to properly describe structure formation. 

The presence of such modes {\it in the numerically evolved fields} implies that any initial fluctuation will inevitably drive one increasingly away from homogeneity and isotropy, already in the linear regime, in contrast with linear perturbation theory around Minkowski space-time. Note that this has nothing to do with spurious growing modes due to numerical error, since the modes we are discussing are physical and present already at the linear analytic level. This poses a problem for setting up a cosmological analogue of the robustness test (see \cite{A2A1, A2A2} for standard testbeds in NR). By the latter we mean testing the code's ability to follow the FLRW space-time, instead of Minkowski, in the presence of a small-amplitude random field perturbation mimicking the numerical error. The importance of such a test is well-established \cite{A2A1, A2A2}, as it constitutes a prerequisite for trusting the stability of simulations in the non-linear regime. It is therefore worth trying to find a way of expressing the system such that the test can be passed. 

Returning to the above example, one can note that while $\de^*$ grows, $\Phi$ does not, because $a^2 \bar{\ro} \sim a^{-1}$ cancels out the growth of $\de^*$ in the Poisson equation. It is then possible to eliminate $\de^*$ from the gravitational dynamical equations, as already noticed by Bardeen \cite{Bardeen}. One can thus evolve only the bounded field $\Phi$ and compute $\de^*$ anytime through the Poisson equation. Here we generalize this idea to the fully non-linear case by noting that a perfect fluid field (PF) can be completely eliminated through the diffeomorphism constraints algebraically. This leaves us with a set of unconstrained evolution equations for the gravitational fields alone which, as is well known, have no growing linear modes in the appropriate gauge. The numerical accuracy of this indirect integration method of the fluid dynamics, e.g. in capturing shock waves, will be addressed in future work \cite{DDM}. 

We illustrate this idea in the ``standard ADM form'' of the Einstein equations and discuss some applications. A particularly original one is that the PF field can be used as an artificial component representing constraint violation, whose dynamics could be chosen such that this violation is damped. This could be very useful in the case where the physical matter content is modeled in the $N$-body approach. Finally, focusing on the case where the PF is the only physical matter content, we consider a specific choice of variables and gauge for which there are no growing linear modes in the evolved fields and corroborate this numerically by successfully passing the robustness test on a FLRW space-time with $p = \La = 0$. This also implies that the scheme solves very accurately the exactly homogeneous and isotropic space-time solution, as we show by an explicit computation.

\section{The general idea} \label{sec:mainidea}

\subsection{Set up}

Let us consider the metric in ADM form
\beq
\ed s^2 = - \al^2 \ed t^2 + \ga_{ij} \( \ed x^i + \be^i \ed t \) \( \ed x^j + \be^j \ed t \) \, ,
\eeq
where $\al$ is the lapse function, $\be^i$ is the shift vector and $\ga_{ij}$ is the induced $3$-metric, so that $n := \al^{-1} \( \pa_t - \be^i \pa_i \)$ is the unit vector normal to the spatial hypersurfaces. The gravitational equations of motion of GR in standard ADM form then read ($8\pi G = c = 1$)
\bea
\( \pa_t - \Lie_{\be} \) \ga_{ij} & = & -2 \al K_{ij} \, , \label{eq:EOMq} \\
\( \pa_t - \Lie_{\be} \) K_{ij} & = & - \al \[ 2 K_{ik} K_j^k - K_{ij} K - R_{ij} + \La \ga_{ij} + S_{ij} - \frac{1}{2}\, \ga_{ij} \( S - E \)  \] - \na_i \na_j \al   \, ,  \label{eq:EOMK}  
\eea
and
\bea
0 & = & E + \frac{1}{2} \( K_{ij} K^{ij} - K^2 - R \) + \La  \, , \label{eq:Hconst} \\
0 & = & P_i - \na_j K_i^j + \na_i K  \, , \label{eq:Mconst}
\eea
where we use $\ga_{ij}$ to displace the indices, $K_{ij}$ is the extrinsic curvature, $\na$ and $R_{ij}$ are the connection and Ricci tensor of $\ga_{ij}$, $\La$ is the cosmological constant and $E$, $P_i$, $S_{ij}$ are the canonical matter energy, momentum and stress densities, respectively. These are the components of the energy-momentum tensor $T_{\mu\nu}$ in the $n$-frame
\bea \label{eq:EPSdef}
E := n^{\mu} n^{\nu} T_{\mu\nu} \, , \hspace{1cm} P^{\mu} := - h^{\mu\nu} n^{\ro} T_{\nu\ro} \, , \hspace{1cm} S^{\mu\nu} := h^{\mu\ro} h^{\nu\si} T_{\ro\si} \, ,
\eea
where
\beq
h_{\mu\nu} := g_{\mu\nu} + n_{\mu} n_{\nu} \, ,
\eeq
is the associated projector. Incidentally, $P^0, S^{0\mu} \equiv 0$, while $P_i := \ga_{ij} P^j$ and $S_{ij} := \ga_{ik} \ga_{jl} S^{kl}$. Assuming the presence of an independently conserved PF, we can split $T_{\mu\nu} \equiv {\cal T}_{\mu\nu} + T^{\rm extra}_{\mu\nu}$, where the curly letters (${\cal E}$, ${\cal P}_i$, ${\cal S}_{ij}$) denote the PF part, while the ``extra" superscript labels the extra matter content. We thus have
\beq \label{eq:TPF}
{\cal T}^{\mu\nu} :=  \( \ro + p \) U^{\mu} U^{\nu} + p\, g^{\mu\nu}  \, ,
\eeq
where $U^{\mu}$ is the fluid's $4$-velocity and $\ro$ and $p$ are the energy density and pressure in the fluid's rest-frame $U$. Using (\ref{eq:EPSdef}) for the PF sector we then get
\bea
{\cal E} & = & W^2 \( \ro + p \) - p \, , \\
{\cal P}^i & = & W^2 \( \ro + p \) v^i  \, , \\
{\cal S}^{ij} & = & W^2 \( \ro + p \) v^i v^j + p \ga^{ij} \equiv \frac{{\cal P}^i {\cal P}^j}{{\cal E} + p} + p \ga^{ij} \, , \label{eq:Hij}
\eea
where
\beq
v^i := \al^{-1} \( \frac{U^i}{U^0} - \frac{n^i}{n^0} \)  \, , \hspace{1cm} W := -n_{\mu} U^{\mu} \equiv \[ 1 - v_i v^i \]^{-1/2} \, ,
\eeq
are the proper time $3$-velocity measured by the canonical observer $n^{\mu}$ and the generalization of the corresponding Lorentz factor, respectively. In particular, we can invert
\beq \label{eq:roE}
\ro = {\cal E} - \frac{{\cal P}_i {\cal P}^i}{{\cal E} + p} \, , \hspace{1cm} v^i = \frac{{\cal P}^i}{{\cal E} + p} \, .
\eeq
Using the above $\ro\,({\cal E}, {\cal P}, p)$ relation, along with the equation of state $p = p(\ro)$, we can express $p$ in terms of ${\cal E}$ and ${\cal P}_i$. For instance, in the case of constant equation of state parameter $w := p/\ro$, we find
\beq
p = \frac{1}{2} \[ \sqrt{\( 1 + w \)^2 {\cal E}^2 - 4w {\cal P}_i {\cal P}^i} - \( 1 - w \) {\cal E} \] \, .
\eeq
Equation (\ref{eq:Hij}) then implies that all PF quantities can be expressed algebraically in terms of ${\cal E}$ and ${\cal P}_i$, i.e. the degrees of freedom of a PF, a fact that is important to keep in mind in what follows.

\subsection{Eliminating the perfect fluid}

In the most common NR schemes, the so-called ``free evolution'' schemes, and in particular in the cosmological simulations performed so far, one solves numerically the evolution equations, i.e. the ones that are differential equations in time. Here this corresponds to solving equations (\ref{eq:EOMq}) and (\ref{eq:EOMK}) for the gravitational sector. For a generic matter sector, one needs to specify the matter action and derive the corresponding equations of motion. However, in the case of PF matter, the fact that there are only four degrees of freedom ${\cal E}$ and ${\cal P}_i$ means that their evolution is entirely determined by the energy-momentum conservation equation 
\beq \label{eq:EMTcons}
\na_{\mu} {\cal T}^{\mu\nu} = 0 \, ,
\eeq
since we have assumed that the PF is minimally coupled and thus conserved independently of $T_{\mu\nu}^{\rm extra}$. Indeed, using the above translations between $n$-frame and $U$-frame components, we can express (\ref{eq:EMTcons}) in terms of ${\cal E}$ and ${\cal P}^i$
\bea
\( \pa_t - \Lie_{\be} \) {\cal E} & = & - \na_i( \al {\cal P}^i ) - {\cal P}^i \na_i \al + \al \( K_{ij} {\cal S}^{ij} + K {\cal E} \)  \, , \label{eq:evolE} \\
\( \pa_t - \Lie_{\be} \) {\cal P}^i & = & - \na_j ( \al {\cal S}^{ij} ) - {\cal E} \na^i \al + \al \( 2 K^i_j {\cal P}^j + K {\cal P}^i \) \, , \label{eq:evolP}
\eea
which form a closed set thanks to (\ref{eq:Hij}) and the equation of state. With the evolution equations being the ones that are solved numerically, the constraint equations (\ref{eq:Hconst}) and (\ref{eq:Mconst}) need only to be solved at the initial data surface and are then monitored at each time-step\footnote{At the analytical level, if these equations are statisfied for the initial data, then they are also satisfied at all times, i.e. they are compatible with the evolution equations, a property that is guaranteed by the structure of a gauge theory.}, thus providing a first and valuable stability check of the simulation. 

Here we propose the opposite choice for dealing with the PF field. Instead of using the four energy-momentum conservation equations (\ref{eq:evolE}) and (\ref{eq:evolP}) to evolve ${\cal E}$ and ${\cal P}_i$ in time, we choose to use the four constraint equations (\ref{eq:Hconst}) and (\ref{eq:Mconst}) to determine ${\cal E}$ and ${\cal P}_i$ algebraically out of the rest of the involved fields at each time-step
\bea
{\cal E} & = & - \frac{1}{2} \( K_{ij} K^{ij} - K^2 - R \) - \La - E^{\rm extra} \, , \label{eq:H} \\
{\cal P}_i & = & \na_j K_i^j - \na_i K - P^{\rm extra}_i \, . \label{eq:Hi}
\eea
With this algebraic substitution, the PF fields are totally eliminated and the resulting system is unconstrained\footnote{Barring gauge-theoretical constraints in the extra matter sector.}. In the standard ADM approach in which we work here, our scheme would therefore correspond to solving (\ref{eq:EOMq}), (\ref{eq:EOMK}) and the evolution equations of the ``extra'' matter fields, where (\ref{eq:H}) and (\ref{eq:Hi}) have been used to eliminate the PF fields. These last two equations can then be used to obtain the PF information out of the fields that are actually been evolved at any time. As a consequence, it is the PF evolution equations (\ref{eq:evolE}) and (\ref{eq:evolP}) that are now redundant, i.e. they are no longer needed for closing the system. In analogy with the free evolution schemes, one could instead monitor these equations to check numerical stability. As already mentioned in the introduction, however, the study of the accuracy of the PF dynamics using this indirect method is postponed to future work. The only check we will perform here is related to the FLRW solution.

From now on let us focus on the case $T^{\rm extra}_{\mu\nu} = 0$ for simplicity. We see that $\ga_{ij}$, $K_{ij}$ parametrize the constraint hypersurface in phase space of the full GR + PF system, i.e. we only have physical states by construction. The numerical constraint violation of free evolution schemes is now translated into an error on the PF quantities ${\cal E}$ and ${\cal P}_i$ and therefore to a violation of energy-momentum conservation $\na_{\mu} {\cal T}^{\mu\nu} \neq 0$, i.e. the relation between two subsequent states is not necessarily physical. It therefore seems that we have only traded one kind of error for another. However, what we did gain is that now the PF fields no longer participate in the robustness test, since they are not evolved in the integration to begin with. They are only physically relevant combinations of the fields that we are actually solving numerically. This way we avoid the growing modes discussed in the introduction (e.g. the matter density contrast), that are present already at the analytical linear level. Since the gravitational perturbations are bounded in time in the appropriate gauge, we are now able to set up a robustness test around a FLRW solution.

\subsection{Fluid degrees of freedom inside the gravitational fields}

Let us finally get a bit more insight into how the gravitational and PF degrees of freedom (DOF) coexist inside $\ga_{ij}$ and $K_{ij}$. To that end, we linearize (\ref{eq:EOMq}) and (\ref{eq:EOMK}) around a flat FLRW background, using (\ref{eq:Hij}), (\ref{eq:H}) and (\ref{eq:Hi}). Being only interested in the spectrum, we look at small enough scales/periods and can thus effectively set the scale factor to $1$. Moreover, we can also integrate out the $K_{ij}$ variables, thus expressing the equations in second-order form. We perform a scalar-vector-tensor (SVT) decomposition
\bea
\al & = & 1 + \psi \, , \\
\be^i & = & \pa^i \chi + \chi^i \, , \\  
\ga_{ij} & = & \( 1 - 2 \si \) \de_{ij} + 2 \pa_i \pa_j h + 2 \pa_{(i} h_{j)} + 2 h_{ij} \, , 
\eea
i.e. all vectors/tensors have zero divergence and trace, and we displace the indices using $\de_{ij}$. We get two wave equations
\beq
\ddot{\si} = \bar{c}_s^2 \pa^2 \si  \, , \hspace{1cm} \ddot{h}_{ij} = \pa^2 h_{ij} \, ,
\eeq
where $c_s^2 := \ed p/ \ed \ro$ is the speed of sound and appears through the perturbation of $p$ in ${\cal S}_{ij}$, using the fact that $\ro = {\cal E}$ at the linear level (\ref{eq:roE}). We see that the gravitational potential $\si$ is now dynamical and carries the sound DOF. In vacuum GR we have that $\bar{c}_s^2 \to 1$, but $\si$ has no DOF because of the Hamiltonian and (the divergence of) the momentum constraints. As for the $h$ and $h_i$ components, it is convenient to integrate in two auxiliary fields $\pi$ and $\pi_i$ (with $\pa_i \pi_i \equiv 0$) in order to express the equations in first-order form
\beq
\dot{h} = \pi + \chi \, , \hspace{1cm} \dot{\pi} =  - \phi + \psi \, , 
\eeq
and
\beq
\dot{h}_i = \pi_i + \chi_i \, , \hspace{1cm} \dot{\pi}_i = 0 \, .
\eeq
The $\pi$ and $\pi_i$ fields are linear combinations of $\ga_{ij}$ and $K_{ij}$ perturbations, whose precise form is irrelevant for our purposes here. Indeed, the relevant feature is the structure of the equations. In particular, we see that there are no DOF in the $(h, \pi)$ couple, because we can trivialize the evolution of both fields by choosing appropriately $\chi$ and $\psi$. As for $(h_i, \pi_i)$, the $h_i$ components can be trivialized with $\chi_i$, so we only have two DOF in $\pi_i$, corresponding to the rotational velocity of the fluid. In vacuum GR the latter are neutralized by the rotational part of the momentum constraint. Thus, the four DOF of the fluid ${\cal E}$ and ${\cal P}_i$ lie in the lower helicity modes of the gravitational field that are now unconstrained. We are therefore basically evolving the PF through the gravitational fields. 

One should also observe that this ``de-constraining'' procedure is only possible in a cosmological context, because we need ${\cal E}+p > 0$ in (\ref{eq:Hij}). This, however, does not mean that we cannot handle local voids ${\cal E} = p = 0$. Indeed, assuming the weak energy condition $p \geq -\ro$, we get from (\ref{eq:roE}) that $({\cal E} + p)^2 \geq {\cal P}_i {\cal P}^i$, so the numerator in ${\cal P}_i {\cal P}_j/({\cal E} + p)$ tends faster to zero than the denominator. One can thus safely set ${\cal S}_{ij} \to \ga_{ij} p$ by hand as soon as ${\cal E}$ goes below some threshold value. 

Finally, note that any generally-covariant theory of gravity can be deconstrained in some analogous manner, simply because ${\cal E}$ and ${\cal P}_i$ will always appear linearly in the diffeomorphism constraints. However, depending on the precise structure of the theory and extra matter content, this might not be enough to get rid of all growing linear modes in the fields that need to be evolved.

\section{A concrete scheme} \label{sec:scheme}

From now on we consider the case where the PF is the only physical matter content in the universe, i.e. $T^{\rm extra}_{\mu\nu} = 0$. 

\subsection{Choice of variables and gauge}

We start by rewriting the unconstrained GR + PF system, i.e. (\ref{eq:EOMq}) and (\ref{eq:EOMK}) with (\ref{eq:Hij}), (\ref{eq:H}) and (\ref{eq:Hi}), in the form which provides a better numerical behavior. We first decompose $\ga_{ij}$ and $K_{ij}$ in the conformal/traceless fashion
\beq
\ga_{ij} \equiv e^{4\ph} \ti{\ga}_{ij} \, , \hspace{1cm} K_{ij} \equiv e^{4\ph} \( \ti{A}_{ij} + \frac{1}{3}\, \ti{\ga}_{ij} K \) \, ,
\eeq
where $\det \ti{\ga}_{ij} \equiv 1$, $\ti{\ga}^{ij} \ti{A}_{ij} \equiv 0$ and all indices are displaced using $\ti{\ga}_{ij}$ from now on. The above separation, also used in the BSSN scheme \cite{BS,SN}, is especially important in cosmology because it distinguishes the fields with a non-trivial background $\ph$, $K$ from the rest $\ti{\ga}_{ij}$, $\ti{A}_{ij}$. We then consider the following slicing
\beq
\al = a^2 e^{-2\ph} \, , \hspace{1cm} a(t) := \( \frac{\int \ed^3 x \, e^{6 \ph}}{\int \ed^3 x} \)^{1/3} \, ,
\eeq
where the integrals are carried over the full spatial slice. On a FLRW space-time, both $a$ and $e^{2\ph}$ (and thus $\al$) reduce to the scale factor, so this is a non-linear generalization of conformal time. This choice is mostly out of convenience, because the linear propagating fluctuations have a constant frequency with respect to that time.

Next, we consider the gauge $\ti{\Ga}^i  := - \pa_j \ti{\ga}^{ij} = 0$, whose conservation $\dot{\ti{\Ga}}^i = 0$ leads to an elliptic equation for $\be^i$. In practice, however, the modified condition $\dot{\ti{\Ga}}^i - \al \la K \ti{\Ga}^i = 0$ with $\la > 0$ helps the convergence of numerical deviations from $\ti{\Ga}^i = 0$, and the detailed equation reads  
\beq \label{eq:beeq}
\ti{\pa}^2 \be^i + \frac{1}{3}\, \ti{\ga}^{ij} \pa_j \pa_k \be^k - 2 \al \( \pa_j \ti{A}^{ij} - 2 \ti{A}^{ij} \pa_j \ph \) - \al \la K \ti{\Ga}^i = 0 \, ,
\eeq
where $\ti{\pa}^2 := \ti{\ga}^{ij} \pa_i \pa_j$ and we have used $\ti{\Ga}^i = 0$ to simplify the $\sim \dot{\ti{\Ga}}^i$ part. The evolution equations (\ref{eq:EOMq}) and (\ref{eq:EOMK}) now read
\bea
\dot{\ph} & = & -  \frac{1}{6} \, \al K + \be^i \pa_i \ph + \frac{1}{6}\, \pa_i \be^i  \, , \label{eq:phidot} \\
\dot{\ti{\ga}}_{ij} & = & -  2 \al \ti{A}_{ij} + \be^k \pa_k \ti{\ga}_{ij} + \ti{\ga}_{kj} \pa_i \be^k + \ti{\ga}_{ik} \pa_j \be^k  - \frac{2}{3}\, \ti{\ga}_{ij} \pa_k \be^k \, , \label{eq:gadot} \\
\dot{K} & = & \frac{\al}{2} \[ K^2 - 3 \La + \frac{3}{2}\, \ti{A}_{ij} \ti{A}^{ij} + e^{-4\ph} \(  \frac{1}{2}\,B + \ti{\ga}^{ij} {\cal S}_{ij}  \) \]  + \be^i \pa_i K \, , \label{eq:Kdot} \\
\dot{\ti{A}}_{ij} & = & \al \[ K \ti{A}_{ij} - 2 \ti{A}_{ik} \ti{A}_j^k + e^{-4\ph} \( B_{ij} - {\cal S}_{ij} \)^{\rm T} \]  + \be^k \pa_k \ti{A}_{ij} + \ti{A}_{kj} \pa_i \be^k + \ti{A}_{ik} \pa_j \be^k - \frac{2}{3}\, \ti{A}_{ij} \pa_k \be^k  \, , \nn \\ \label{eq:Adot} 
\eea
where
\bea
B_{ij} & := & - \frac{1}{2}\, \ti{\pa}^2 \ti{\ga}_{ij} + \ti{\ga}^{kl} \ti{\ga}^{mn} \( 2 \ti{\Ga}_{km(i} \ti{\Ga}_{j)ln} + \ti{\Ga}_{kmi} \ti{\Ga}_{lnj} \) - 8 \pa_i \ph \pa_j \ph \, , \\
B & := & \ti{\ga}^{ij} B_{ij} = \ti{\ga}^{ij} \( \ti{\Ga}^k_{li} \ti{\Ga}^l_{kj} - 8 \pa_i \ph \pa_j \ph \)  \, , \label{eq:B} \\
{\cal S}_{ij} & \equiv & e^{4\ph} \ti{\ga}_{ij} p + \frac{{\cal P}_i {\cal P}_j}{{\cal E} + p} \, , \\
{\cal E} & = &  - \frac{1}{2}\, \ti{A}_{ij} \ti{A}^{ij} + \frac{1}{3}\, K^2 - \La + \frac{1}{2}\, e^{-4\ph}  \( B - 8 \ti{\pa}^2 \ph \)   \, , \\
{\cal P}_i & = & \pa_j \ti{A}_i^j - \ti{A}^j_k \ti{\Ga}^k_{ji} + 6 \ti{A}_i^j \pa_j \ph - \frac{2}{3}\, \pa_i K  \, , 
\eea
``T'' denotes the traceless part and $\ti{\Ga}^i_{jk} \equiv \ti{\ga}^{il} \ti{\Ga}_{ljk}$ are the Christoffel symbols of $\ti{\ga}_{ij}$. Thanks to $\al \sim e^{-2\ph}$ and $\ti{\Ga}^i = 0$, the only double spatial derivatives that appear are conformal Laplacians $\ti{\pa}^2$, a ``hyperbolic'' characteristic which could be responsible for the good numerical performance. 

Here it is important to stress that, although we do use the conformal/traceless decomposition, which is one of the defining features of the BSSN scheme, there is another crucial feature of BSSN which cannot be implemented here and therefore clearly distinguishes the two approaches. Indeed, in BSSN one evolves independently $\ti{\Ga}^i$, considering its definition as an extra constraint equation, and one uses the momentum constraint to alter its evolution equation. This last manipulation is ``essential for the numerical stability of the system'' \cite{BS}, because it is needed to make the system strongly hyperbolic \cite{AABSS, SCPT}. In our case, the fact that the constraint equations have been used to eliminate part of the fields means that we can no longer use them at all, since this would reintroduce the said fields in the equations. We are therefore far from the defining structure of the BSSN scheme on which many of its merits are based.

\subsection{Absence of growing linear modes} \label{sec:nogrow}

Let us now show that there are no growing linear modes, at the analytical level, in the evolved fields in this gauge for the simpler case of constant equation of state parameter $w := p/\ro$. We thus linearize around FLRW and perform a SVT decomposition
\bea
\ph & = & \frac{1}{2} \[ \log a + h \] \, , \\
\ti{\ga}_{ij} & = & \de_{ij} + 2 \[ \( \pa_i \pa_j - \frac{1}{3}\, \de_{ij} \pa^2 \) \ti{h} + \pa_{(i} \ti{h}_{j)} + \ti{h}_{ij} \] \, , ~~~ \\
K & = & - 3 H \[ 1 + \ka \] \, ,  \label{eq:Kpert} \\
\ti{A}_{ij} & = & - H \[  \( \pa_i \pa_j - \frac{1}{3}\, \de_{ij} \pa^2 \) \ti{\ka} + \pa_{(i} \ti{\ka}_{j)} + \ti{\ka}_{ij} \] \, , \label{eq:Aijpert} \\
\al & = & a \( 1 - h \) \, ,   \\
\be^i & = & - a H \[ \( \pa^i \ti{\ka} + \ti{\ka}^i \) + \la \( \pa^i \ti{h} + \ti{h}^i \) \] \, , 
\eea
where $a(t)$ is the scale factor and $H(t) := (a^{-1} \pa_t) \log a$ is the physical Hubble parameter in conformal time. If we further set $\ti{\Ga}^i = 0$, then $\ti{h}$ and $\ti{h}_i$ vanish. However, we keep these components to verify that the system maintains their amplitude close to numerical error. Going to Fourier space and choosing the time variable $x := \log a$ the equations become
\beq
\pa_x S = M_s S \, , \hspace{0.5cm} \pa_x V = M_v V \, , \hspace{0.5cm} \pa_x T = M_t T \, ,  
\eeq
for the scalar $S := ( h, \ka, \ti{h}, \ti{\ka} )$, vector $V := ( \ti{h}_i, \ti{\ka}_i )$ and tensor $T := ( \ti{h}_{ij}, \ti{\ka}_{ij} )$ multiplets with
\beq
M_s = \( \begin{array}{cccc} -1 & 1 & \la k^2/3 & k^2/3 \\  - w\ti{k}^2 & - X & 0 & 0 \\ 0 & 0 & -\la & 0 \\ 0 & 0 &  -\ti{k}^2 & 3w - X \end{array} \) \, , \hspace{0.5cm} M_v = \( \begin{array}{cc} -\la & 0 \\ -\ti{k}^2 & 3w - X \end{array} \) \, , \hspace{0.5cm} M_t = \( \begin{array}{cc} 0 & 1 \\  -\ti{k}^2 & 3w - X \end{array} \)    \, ,  
\eeq
where $\ti{k} := k/(a H)$, $X := (3/2) \( 1+w \) \( 1 + \Om_{\La} \) $ and $\Om_{\La} := \La/(3 H^2) < 1$. All eigenvalues have negative real part for the typical cases of interest $0 \leq w \leq 1/3$, $\Om_{\La} \geq 0$ and $\la > 0$. Moreover, the determinant of the eigenvector matrices is generically non-zero, so that $M_s$, $M_v$ and $M_t$ are diagonalizable and one can thus trust the spectrum to deduce the behavior.

\section{Numerical tests}

In this section we specialize to the case $p = 0$ and $\La = 0$, i.e. a universe filled with only pressureless matter. We have developed our code using the \lf\ library \cite{DHB,LF_WEB}.  \lf\ is a C++ library designed to simplify writing parallel codes for solving partial differential equations on rectilinear meshes, developed for application to classical field theories in particle physics and cosmology. The problem is discretized on a regular Cartesian mesh. Spatial derivatives are computed using finite differences with 3,5, or 7 point stencil. In this paper the presented results are computed using a 5 points stencil. The time integration of (\ref{eq:phidot}), (\ref{eq:gadot}), (\ref{eq:Kdot}) and (\ref{eq:Adot}) is performed using a fourth-order Runge-Kutta method (RK4) without any artificial dissipation method. The algebraic constraints $\det \ti{\ga}_{ij} \equiv 1$, $\ti{\ga}^{ij} \ti{A}_{ij} \equiv 0$ are imposed by hand \cite{A2A2,Aletal,ABCBRP}  at each intermediary RK4 step. The elliptic equation imposing our gauge condition (\ref{eq:beeq}) is solved using a relaxation method \cite{Okawa} at every time-step with $\la = 100$ and the $\be^i$ from the previous time-step as initial conditions. Finally, from now on all distances are given in ${\rm Mpc}/h$ units, where here $h$ is the dimensionless Hubble constant, i.e. $H_0 \equiv 100 h \, {\rm km\, s^{-1}\, Mpc^{-1}}$.

\subsection{The FLRW solution}

\begin{figure}[h!]
\includegraphics[width=0.9\columnwidth]{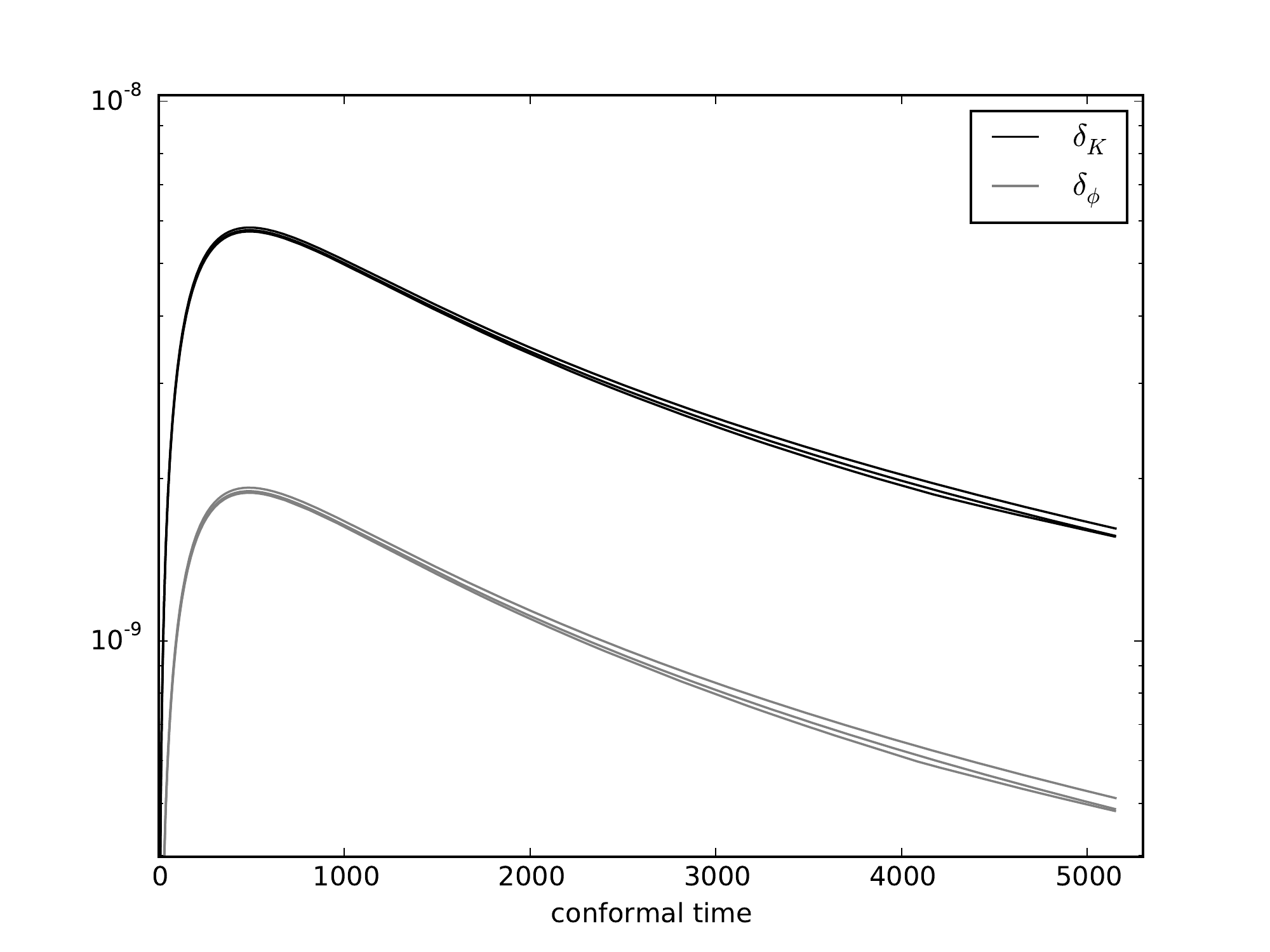} 
\caption{Relative errors for the FLRW solution. We plot the normalized deviations $N^4\de_{\ph}$ (grey lines) and $N^4 \de_K$ (black lines) for three resolutions $\De t = 3.2/N$, where $N \in \{ 1,2,4 \}$. The magnitude of the curves decreases with increasing $N$.}
\label{fig:FLRW}
\end{figure}

We first consider integrating the homogeneous and isotropic solution with zero spatial curvature. For the precise FLRW solution we will consider, the non-trivial fields are given by  
\bea
\ph_{\rm FLRW}(t) & = & \frac{1}{2}\, \log a_{\rm FLRW}(t) = \log \[ \frac{t}{t_0} + \sqrt{a_i} \( 1 - \frac{t}{t_0} \) \]   \, , \\
K_{\rm FLRW}(t) & = & - 3 H_{\rm FLRW}(t) = - \frac{6 \( 1 - \sqrt{a_i} \)}{t + \sqrt{a_i} \( t_0 - t \)} \, , 
\eea
where the ``$i$" and ``$0$" subscripts denote the initial and final times, respectively, and in particular we have chosen $t_i = 0$. Incidentally, the initial and final values of the scale factor are $a_i$ and $a_0$, and here we have $a_0 = 1$. The value of $t_0$ is given by
\beq
t_0 = 2 H_0^{-1} \( 1 - \sqrt{a_i} \) \, ,
\eeq
where $H_0$ is the final physical Hubble parameter and we set $H_0^{-1} = 3000$. Finally, we choose $a_i = 1/50$, meaning that we evolve from redshift 49 to 0, and leading to $t_0 \approx 5150$. The initial conditions of our integration being
\beq
\ph(0, \vec{x}) = \ph_{\rm FLRW}(0)  \, , \hspace{1cm} K(0, \vec{x}) = K_{\rm FLRW}(0) \, , \hspace{1cm} \ti{\ga}_{ij}(0, \vec{x}) = \de_{ij} \, , \hspace{1cm} \ti{A}_{ij}(0,\vec{x}) = 0 \, ,
\eeq
the numerical evolution preserves both the lack of $\vec{x}$-dependence as well as the trivial values of $\ti{\ga}_{ij}$ and $\ti{A}_{ij}$. Thus, the non-trivial numerical information is solely $\ph(t)$ and $K(t)$, whose values we pick at some arbitrary point $\vec{x}$ on the grid. More precisely, we are interested in the relative errors with respect to the analytical solution
\beq
\de_{\ph}(t) := |\ph(t) - \ph_{\rm FLRW}(t) | \, , \hspace{1cm} \de_K(t) := \left| \frac{K(t) - K_{\rm FLRW}(t)}{K_{\rm FLRW}(t)} \right| \, ,
\eeq
where $\de_{\ph}$ is indeed ``relative" since the actual gravitational field is $\ga_{ij} \sim e^{4\ph}$. On the PF side, the only non-trivial quantity here is the energy density ${\cal E}(t)$, since $p \equiv 0$, and through (\ref{eq:H}) we get ${\cal E} = K^2/3$. Therefore, the relative error on the energy density is of the same order
\beq 
\de_{\cal E}(t) := \frac{|{\cal E}(t) - {\cal E}_{\rm FLRW}(t)|}{{\cal E}_{\rm FLRW}(t)} \approx 2 \de_K(t)   \, ,
\eeq
by construction. We perform the computation for three time resolutions $\De t = 3.2 / N$, where $N \in \{ 1, 2, 4 \}$, and plot the normalized deviations $N^4 \de_{\ph}$ and $N^4\de_K$ in figure \ref{fig:FLRW}. We first note that the relative errors are bounded in time and exhibit the same behavior for all resolutions. The fact that the curves decrease in magnitude as $N$ increases means that we have achieved fourth-order global convergence, as expected with the RK4 method.

\subsection{The FLRW robustness test}

In section \ref{sec:nogrow} we demonstrated the absence of growing linear modes around the FLRW solution in theory, so the remaining non-trivial requirement is that the fully non-linear system is able to respect that condition numerically. We therefore consider the FLRW initial data plus random perturbation fields
\bea
\ph(0, \vec{x}) & = & \ph_{\rm FLRW}(0) + \vep^{\ph}(\vec{x}) \, , \\
K(0, \vec{x}) & = & K_{\rm FLRW}(0) \[ 1 + \vep^K(\vec{x}) \] \, , \\
\ti{\ga}_{ij}(0, \vec{x}) & = & \frac{\de_{ij} + \vep^{\ga}_{ij}(\vec{x})}{\[ \det \( \de_{kl} + \vep^{\ga}_{kl}(\vec{x}) \) \]^{1/3}} \, , \\
\ti{A}_{ij}(0,\vec{x}) & = & \[ \de_i^k \de_j^l - \frac{1}{3}\, \ti{\ga}_{ij}(0,\vec{x})\, \ti{\ga}^{kl}(0,\vec{x}) \] \vep^{A}_{kl}(\vec{x}) \, ,
\eea
where the entries of $\vep^{\ph, K}$ and $\vep^{\ga,A}_{ij} \equiv \vep^{\ga,A}_{ji}$ are drawn with a uniform distribution of amplitude $\vep$ independently at each space point and for each field component, using the GSL random generator with a different seed for each MPI process. We set periodic spatial boundary conditions and consider a fixed comoving box size of $L = 1024$, so that we start outside the horizon at early times and end up inside at late times. We perform the test for three resolutions, $64^3$, $128^3$ and $256^3$ grid points. We pick $\vep = 10^{-12} \De x^2$, where $\De x \in \{ 16, 8, 4 \}$ is the lattice spacing, so that the matter density contrast $\de \sim \vep\, \pa^2 \ph$ is of the same order for all resolutions. It reaches a maximum value of $\de \sim \Ord(10^{-4})$ at $t_0$, so the linear regime is maintained. Finally the Courant factor is set to 10, i.e. $\De t = \De x/10$, since the speed of light is unity. In terms of $\De t$, our three resolutions therefore correspond to $\De t \in \{ 1.6, 0.8, 0.4 \}$.

\begin{figure}[h!]
\includegraphics[width=0.9\columnwidth]{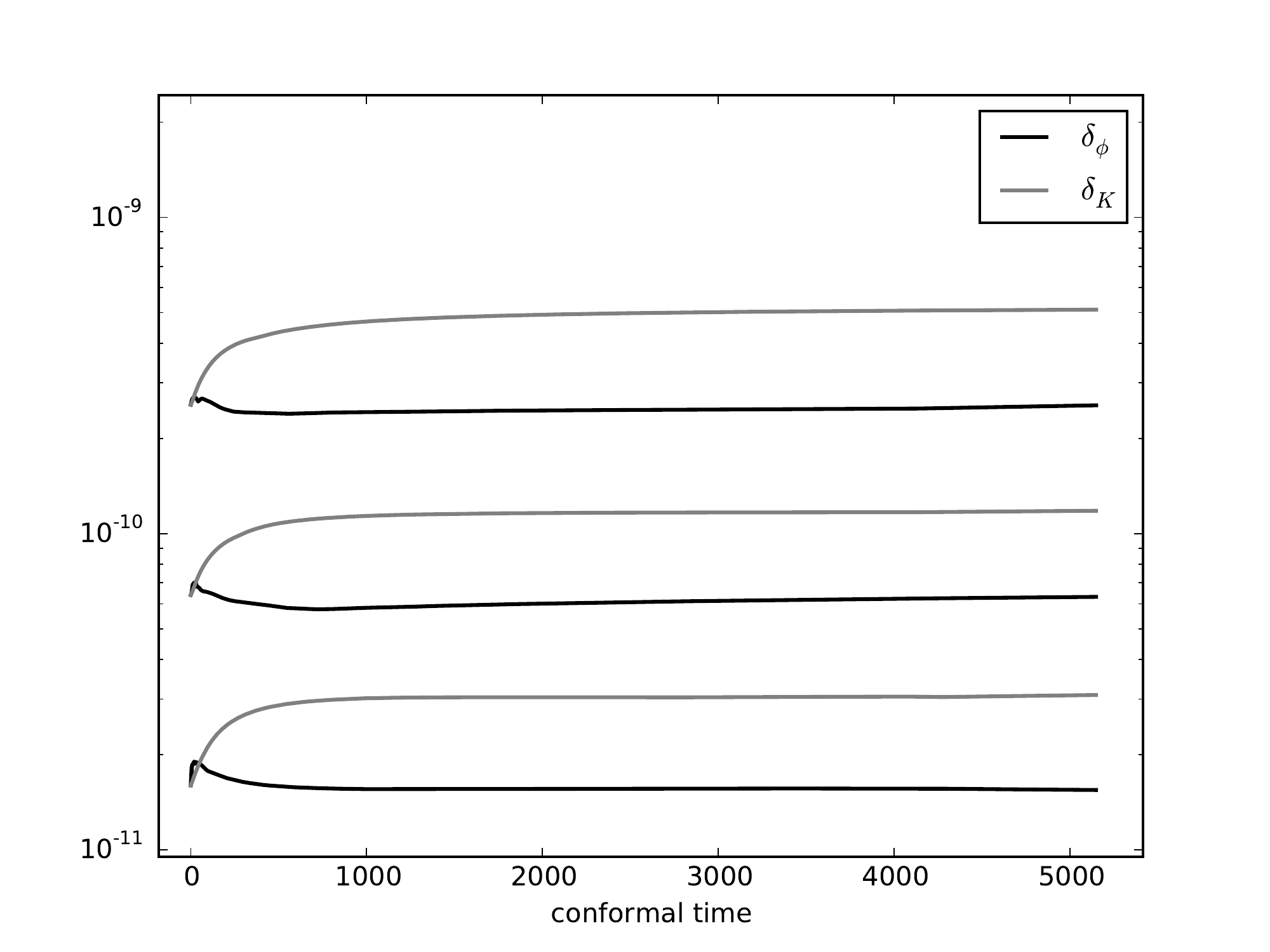} 
\caption{Robustness test around the FLRW solution. We plot $\de_{\ph}, \de_K$ for three resolutions: $64^3$ (top line), $128^3$ (middle line) and $256^3$ (bottom line) grid points.}
\label{fig:robust1}
\end{figure}

\begin{figure}[h!]
\includegraphics[width=0.9\columnwidth]{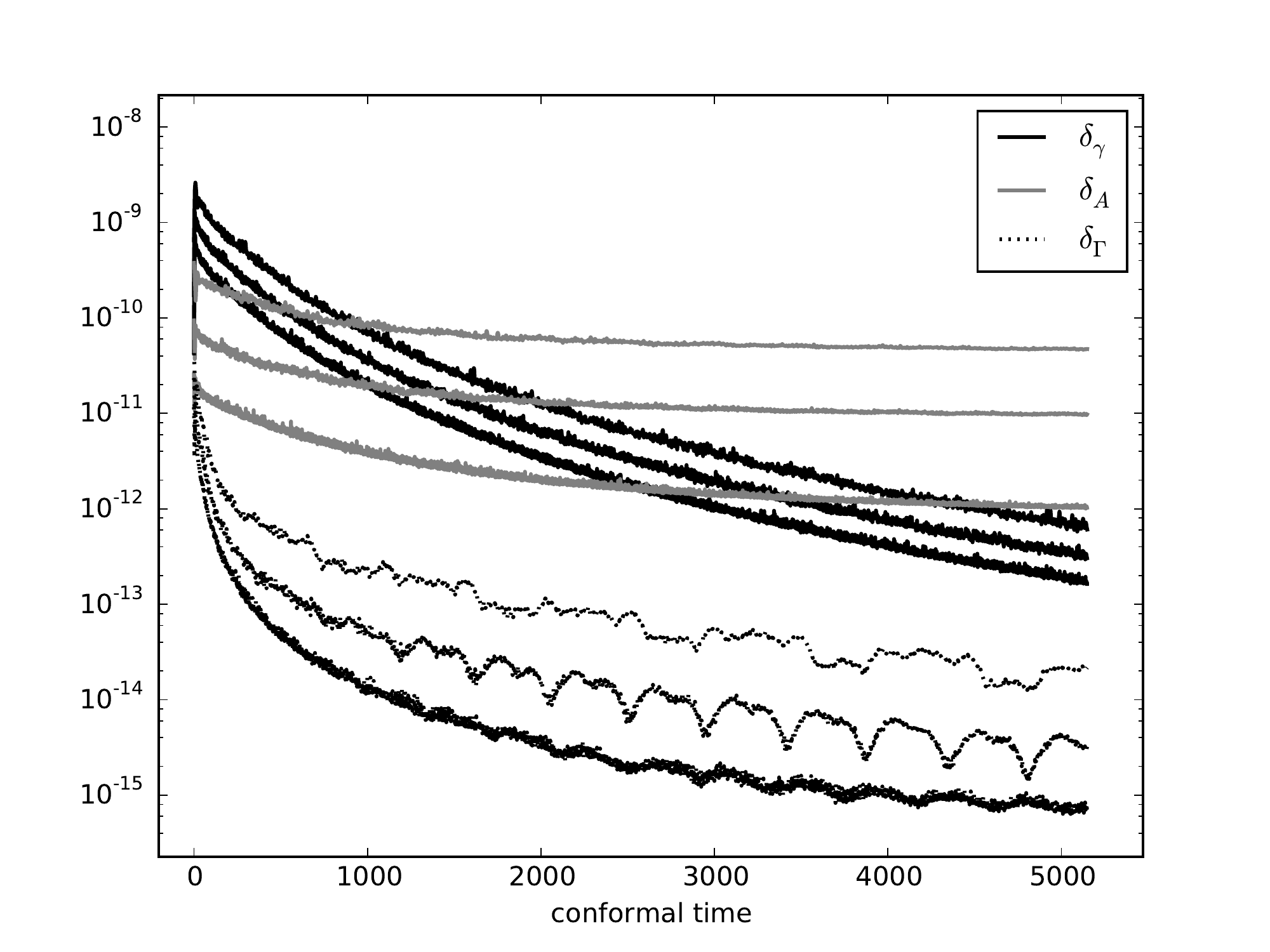}
\caption{Robustness test around the FLRW solution. We plot $\de_{\ga}, \de_A, \de_{\Ga}$ for three resolutions: $64^3$ (top line), $128^3$ (middle line) and $256^3$ (bottom line) grid points.}
\label{fig:robust2}
\end{figure}

We next define the following measures of deviation from the analytical FLRW solution
\beq
\de_{\ph}(t) := \max_{\vec{x}} |\ph(t,\vec{x}) - \ph_{\rm FLRW}(t) | \, , \hspace{1cm} \de_K(t) := \max_{\vec{x}} \left| \frac{K(t,\vec{x}) - K_{\rm FLRW}(t)}{K_{\rm FLRW}(t)} \right| \, , 
\eeq
\beq
\de_{\ga}(t) := \max_{\vec{x}} \sqrt{\De \ti{\ga}_{ij}(t,\vec{x})\, \De \ti{\ga}^{ij}(t,\vec{x})} \, , \hspace{1cm} \de_{A}(t) := \max_{\vec{x}}  \sqrt{\ti{A}_{ij}(t,\vec{x})\, \ti{A}^{ij}(t,\vec{x})}  \, , 
\eeq
where $\De \ti{\ga}_{ij} := \ti{\ga}_{ij} - \de_{ij}$, and the gauge violation
\beq
\de_{\Ga}(t) := \max_{\vec{x}}  \sqrt{\ti{\Ga}_i(t,\vec{x})\, \ti{\Ga}^i(t,\vec{x})}  \, ,
\eeq 
respectively. Note that, in the case of $\ti{\ga}_{ij}$ and $\ti{A}_{ij}$, we can only consider absolute deviations since their FLRW values are trivial. In figures \ref{fig:robust1} and \ref{fig:robust2} we plot the evolution of these quantities for all three resolutions. We see that the perturbations are indeed bounded in time and, most importantly, that we have convergence under resolution increase. More precisely, here it is not the amplitude of the deviation that is relevant, because we are decreasing its initial value by hand  as we increase the resolution $\vep \sim \De x^2$. Rather, what matters is that the behavior of the deviation remains qualitatively the same for all three resolutions. As for the violation of the gauge condition $\de_{\Ga}$, it is well controlled in that it decreases with time and its amplitude is several orders of magnitude below the one of $\de_{\ga}$. Finally, as in the previous subsection, here too the PF quantities have the same level of accuracy as the gravitational ones, since they are determined algebraically by the latter.

\section{Discussion}

In this paper we have proposed a scheme which is able to evolve the FLRW solution stably under small random inhomogeneous perturbations. This does not mean that the more common ``free" schemes, that are currently used for cosmological simulations, are numerically unstable. Rather, these schemes simply cannot pass a FLRW robustness test because of the growing linear modes in the matter sector that are present already at the analytical level. However, we believe that being able to pass that test is valuable for trusting the simulations in the non-linear regime. This is especially true for cosmology, because the relativistic/non-linear effects one wishes to capture by considering NR simulations are indeed very small and could therefore easily be polluted by noise.

Now since our scheme must contain a PF field in order to ``de-constrain'' itself, we should think about its possible interpretations. At a purely theoretical level, there are interesting subjects one can investigate about GR with a globally distributed perfect fluid, such as the backreaction of fluctuations on the evolution of averaged quantities for instance \cite{GMS, BB, MLP}. In the case of structure formation simulations, the shell-crossing phenomenon of cold dark matter can only be captured through an $N$-body description, so a natural choice would be to associate the PF field with baryons. Unfortunately, the complexity of baryonic physics is hardly captured by a PF approximation at scales below $\Ord(1)$ Mpc \cite{vDSBV, vDSCBV, Setal, ST}, so this depends on the scale or the degree of realism one wishes to reach. 

Finally, one can also consider no PF field at all, that is, to avoid giving it a physical interpretation and keep its density intentionally close to zero (with the aforementioned numerical trick ${\cal S}_{ij} \to \ga_{ij} p$). This then provides an elegant mechanism for enforcing constraint conservation on the physical components (gravity + extra matter). Indeed, starting with no ``fluid" ${\cal E}, {\cal P}_i = 0$, i.e. on the constraint surface, and considering an equation of state $p = \ro/3$, any local constraint violating fluctuation will be diluted because of pressure, since small-amplitude radiation does not cluster. The tendency of the equations should thus be to hold the system around the constraint surface, but this remains to be explored numerically. Of particular interest would be the situation where one considers only an $N$-body sector as the physical matter content and adds the artificial PF to control the constraints.

\acknowledgments

We are very grateful to Julian Adamek, Eloisa Bentivegna, Ruth Durrer, Pierre Fleury, Martin Kunz, Luis Lehner, James Mertens, Joachim Stadel and the referees for useful discussions, comments, corrections and suggestions. This work has been supported by several grants of the Swiss National Science Foundation (for all authors), by the Tomalla foundation (for EM) and by a grant from the Swiss National Supercomputing Centre (CSCS) under project ID d55. The simulations have been performed on the Baobab cluster of the University of Geneva, on Piz Daint of the CSCS and on the Cambridge COSMOS SMP system (part of the STFC DiRAC HPC Facility supported by BIS NeI capital grant ST/J005673/1 and STFC grants ST/H008586/1, ST/K00333X/1). YD has been supported by the mobility grant of Heidelberg University, funded by the Excellence Initiative in Germany, and wishes to thank Heidelberg University and the Perimeter Institute for their hospitality during part of the production of this work.

\end{document}

%% file: mydefs.tex
\newcommand{\beq}{\begin{equation}}
\newcommand{\eeq}{\end{equation}}
\newcommand{\bea}{\begin{eqnarray}}
\newcommand{\eea}{\end{eqnarray}}
\newcommand{\ben}{\begin{enumerate}}
\newcommand{\een}{\end{enumerate}}

\newcommand{\pa}{\partial}

\newcommand{\na}{\nabla}
\newcommand{\ed}{{\rm d}}

\newcommand{\ti}{\tilde}

\renewcommand\({\left(}
\renewcommand\){\right)}
\renewcommand\[{\left[}
\renewcommand\]{\right]}

\newcommand{\nn}{\nonumber}

\newcommand{\al}{\alpha}
\newcommand{\be}{\beta}
\newcommand{\ga}{\gamma}
\newcommand{\Ga}{\Gamma}
\newcommand{\de}{\delta}
\newcommand{\De}{\Delta}

\newcommand{\vep}{\varepsilon}

\newcommand{\ka}{\kappa}
\newcommand{\la}{\lambda}
\newcommand{\La}{\Lambda}
\newcommand{\ro}{\rho}
\newcommand{\si}{\sigma}

\newcommand{\ph}{\phi}

\newcommand{\Om}{\Omega}

\newcommand{\Lie}{{\cal L}}
\newcommand{\Ord}{{\cal O}}